\documentclass[journal]{IEEEtran}

\usepackage{graphicx}
\usepackage{caption}
\usepackage{subcaption}
\usepackage{amsmath}
\usepackage{amsfonts}
\usepackage{amssymb}
\usepackage{cite}
\usepackage{booktabs}

\begin{document}

\title{Robust Modulation Technique for Diffusion-based Molecular Communication in Nanonetworks}


\author{Subodh Pudasaini, Seokjooo Shin and Kyung Sup Kwak\\

\thanks{S. Pudasaini and K. S. Kwak are with UWB Wireless Communications Research Center, Inha University, Incheon, Korea.}
\thanks{S. Shin is with Department of Computer Engineering, Chosun University, Gwangju, Korea.}
\thanks{This work was supported by the National Research Foundation of Korea grant funded by the Korean government (MEST 2010-0018116).}
}

%



\maketitle

\begin{abstract}

Diffusion-based molecular communication over nanonetworks is an emerging communication paradigm that enables nanomachines to communicate by using molecules as the information carrier. For such a communication paradigm, Concentration Shift Keying (CSK) has been considered as one of the most promising techniques for modulating information symbols, owing to its inherent simplicity and practicality. CSK modulated subsequent information symbols, however, may interfere with each other due to the random amount of time that molecules of each modulated symbols take to reach the receiver nanomachine. To alleviate Inter Symbol Interference (ISI) problem associated with CSK, we propose a new modulation technique called Zebra-CSK. The proposed Zebra-CSK adds \textit{inhibitor molecules} in CSK-modulated molecular signal to selectively suppress ISI causing molecules. Numerical results from our newly developed probabilistic analytical model show that Zebra-CSK not only enhances capacity of the molecular channel but also reduces symbol error probability observed at the receiver nanomachine.
\end{abstract}


\section{Introduction}

Recent advancements in nanotechnology have offered several practical approaches for the realization of physical, biological or even hybrid nanomachines \cite{akyildiz}. Typically, such nanomachines are a few tens nanometers in size and are able to perform simple tasks such as computing, storage, sensing and actuation. Nanomachines with communication capabilities can be interconnected to form a nanonetwork through which complex tasks that may be relevant and necessary for realizing different biomedical, environmental and industrial applications can be executed in a collaborative manner.

One of the most promising paradigms to interconnect nanomachines, especially the biological and bio-physical hybrid ones, to set up a nanonetwork is molecular communication. Molecular communication involves the transmission of information encoded using molecules that physically travel from a transmitter nanomachine to a receiver nanomachine. Several types of molecular transport mechanisms have been studied so far. Broadly, they involve either passive transport of molecules utilizing free particle diffusion dynamics or active transport of molecules using bacterial chemotaxis and molecular motors that generate motion \cite{nakano}. In this work, we consider the former diffusion-based passive molecular transport mechanism.

Recently, Kuran et al.\ proposed two different modulation schemes for realizing molecular communication via diffusion: Concentration Shift Keying (CSK) and Molecule Shift Keying (MoSK) \cite{kuran1}. They are analogous to amplitude shift keying and frequency shift keying, respectively, which have been popularly used in electromagnetic communication for many decades. CSK uses different concentrations (or simply number) of molecules to uniquely represent different information symbols while MoSK uses different types of molecules for such purpose. For the transmission of $s$ information bits in one MoSK-modulated symbol, $2^s$ different types of molecules are required. Hence, the complexity of transmitter and receiver nanomachines increases as $s$ increases and thus MoSK may not be practical. 




Consider serial transmission of CSK-modulated information symbols over a time-slotted diffusion channel as described in Section II. In such a diffusion channel, for a transmitted information symbol, molecular concentration (or alternately number of messenger molecules) observed at the receiver nanomachine is initially (i.e. at the start of a slot) zero and it quickly increases until reaching its maximum. Then, the molecular concentration slowly decreases over time resulting in a long-tailed molecular concentration signal that may stretches over several slot durations. Messenger molecules belonging to the tail of the molecular concentration signal of \textit{previous} information symbols thus act as the source of interference to the molecular concentration signal of the \textit{current} information symbol. Such interference is called Inter Symbol Interference (ISI). In \cite{pierobon} and \cite{kuran2}, authors have shown that ISI significantly deteriorates the decoding performance of the receiver nanomachine. Motivated with this problem, in this letter, we present a robust modulation scheme, Zebra-CSK, that attempts to reduce ISI as much as possible.

\section{System Model}

We consider a molecular communication system consisting of a pair of transmitter and receiver nanomachines. The transmitter nanomachine is fixed at the origin of an unbounded three-dimensional stationary fluidic environment while the receiver nanomachine is $d$ distance apart from the transmitter nanomachine.

Furthermore, we consider the molecular communication system is time-slotted with a slot duration of $T_s$, and the transmitter and receiver nanomachines are perfectly synchronized. At the transmitter nanomachine, a binary CSK modulation is employed; impulse of $n$ molecules are released at the start of the slot for binary symbol $1$  while no molecules are released for binary symbol $0$. The encoded information symbols will be conveyed to the receiver nanomachine through molecular diffusion. Once a molecule arrives at the receiver nanomachine, it will be removed from the communication medium. We assume that all molecules are homogeneous and diffuses independently with respect to other molecules with a common diffusion coefficient found using the Einstein relation

\begin{equation}
D=\frac{K_B T}{6 \pi \eta r},
\end{equation}
\noindent where $K_B$ is the Boltzmann constant ($K_B =1.38 \times 10^{-23}$), $T$ is the temperature in kelvin, $\eta$ is the viscosity of the fluidic medium, and $r$ is the common radius of the molecules.

The receiver nanomachine counts the total number of received molecules (denoted by a random variable $N$) at the end of each time slot and applies the following decision rule to determine the transmitted binary symbol:

\begin{equation}
\label{eqnDR}
N \mathop{\gtrless}_{0}^{1} \lambda,
\end{equation}

\noindent where $\lambda$ is a pre-specified threshold. 

\section{The Proposed Modulation Technique}




Following two classes of molecules are used in the proposed Zebra-CSK: (i) messenger molecules for encoding information symbols, and (ii) inhibitor molecules for supressing residual messenger molecules from the previous information symbol.

\begin{figure}[b]
\centering
\label{PE1}
\includegraphics[scale = .5]{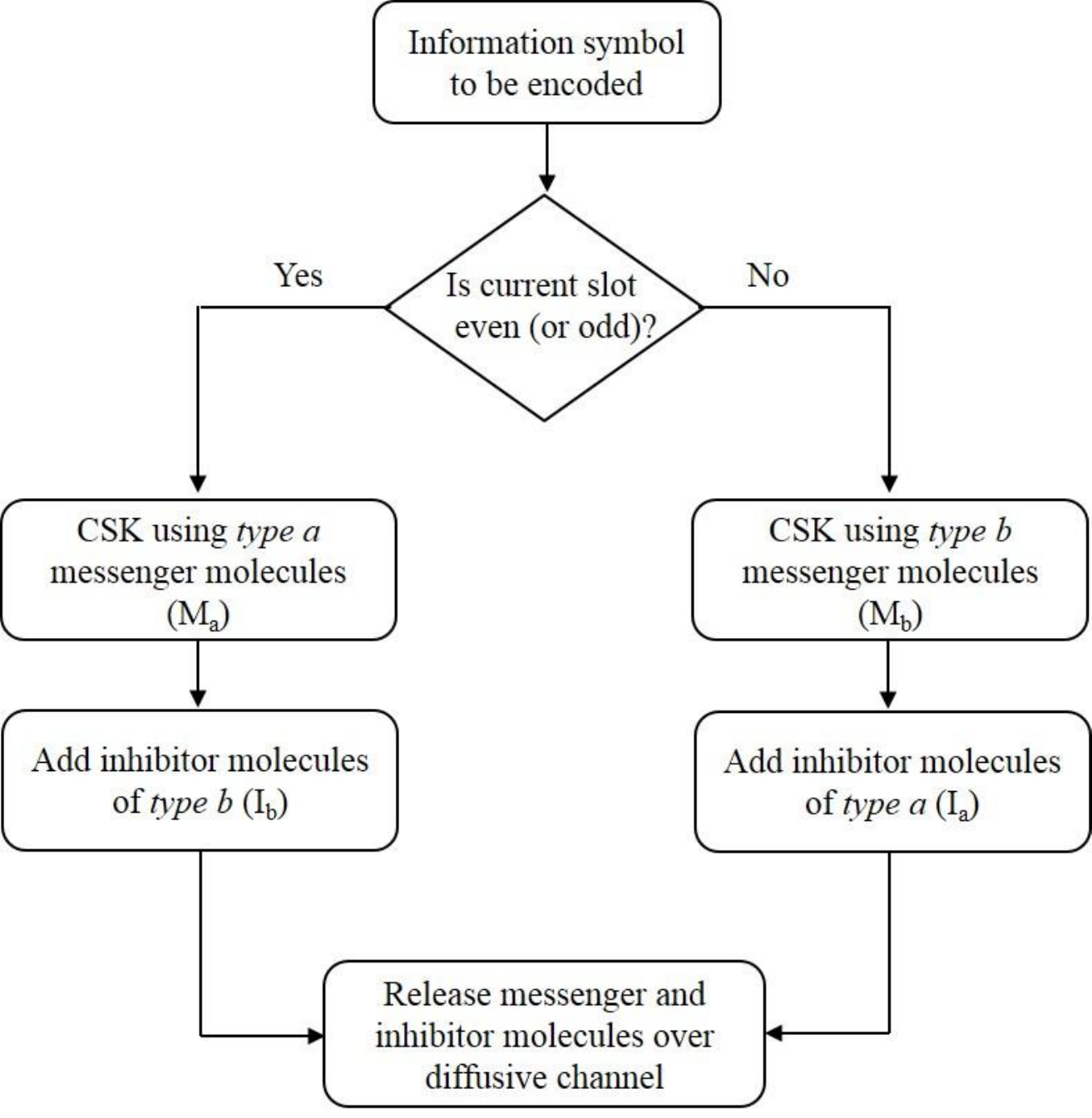}
\caption{\small High-level flowchart of the proposed Zebra-CSK. }
\end{figure}

Fig. 1 depicts the high-level flowchart of Zebra-CSK. In Zebra-CSK, as the name suggests, types of messenger molecules in subsequent information symbols are altered from \emph{type a} messenger molecules (denoted as M$_a$) to \emph{type b} messenger molecules (denoted as M$_b$), or vice versa, while the information encoding mechanism is similar with that of the conventional CSK. Furthermore, each type of messenger molecules are accompanied by inhibitor molecules of the other type. In other words, M$_a$ molecules are accompanied with the inhibitor of M$_b$ molecules (denoted as I$_b$) while M$_b$ molecules are accompanied with the inhibitor of M$_a$ molecules (denoted as I$_b$).

\begin{figure}[t]
\centering
\label{PE1}
\includegraphics[scale = .48]{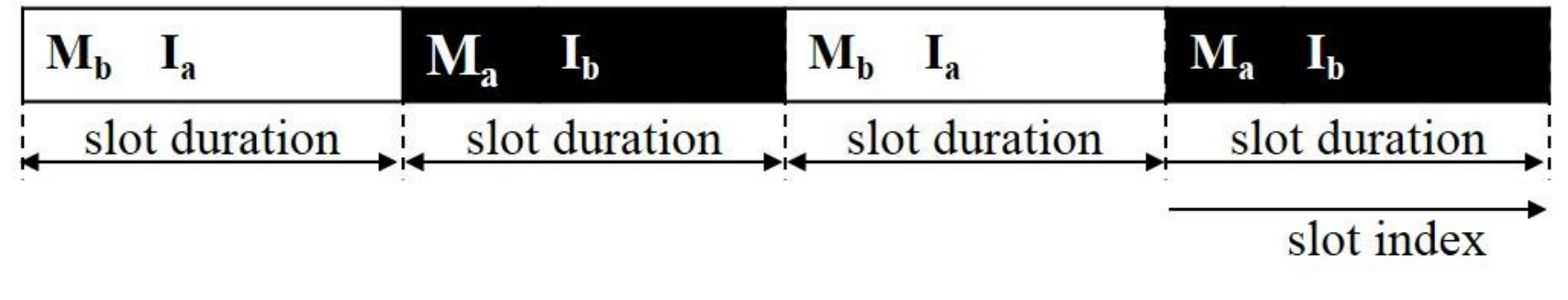}
\caption{\small Temporal molecular-type alternation of both inhibitor and messenger molecules in the proposed Zebra-CSK. }
\end{figure}

Fig. 2 illustrates temporal molecular-type alternation of both messenger and inhibitor molecules in the proposed Zebra-CSK. In a given slot duration, when one type of messenger molecules are released along with  inhibitor of the other type messenger molecules, the late-arriving residual messenger molecules from the previous symbol (which are of the other type) are acted upon by the inhibitor molecules. As a consequence, ISI is reduced. However, it is noteworthy to mention that magnitude of the reduction in ISI depends on the efficiency of the used inhibitor molecules.




\section{Performance Analysis}

\subsection{Analytical Model}
The impulse of messenger molecules released from the transmitter nanomachine spread out through the fluid medium via diffusion and hence the random motion of the messenger molecules can be represented as Brownian motion. Following the well developed literature on analysis of Brownian motion, probability density function of the time a messenger molecule requires to reach the receiver nanomachine (referred to as absorbance time of Brownian particle in the literature) can be written as
\begin{equation}
f(t)=
\begin{cases}
0, & t = 0  \\
 \frac{d}{\sqrt{4\pi D t^3}}\cdot \exp(-\frac{d^2}{4Dt}), & t > 0.
\end{cases}
\end{equation}
\noindent Utilizing (3), the probability that a molecule reaches the receiver nanomachine within a current slot of $T_s$ can be obtained as
\begin{equation}
p_{c}=\int_0^{T_s} f(t)dt.
\end{equation}
\noindent Define $Z_k$ to be a binary random variable indicating whether a $k$-th molecule among $n$ molecules released by the transmitter nanomachine  at the beginning of a slot arrives at the receiver nanomachine before the slot ends, i.e. $Z_k=1$ if the molecule arrives within $T_s$ and $Z_k=0$ otherwise. Thus, considering $Z_k$, we have
\begin{equation}
P(N_c = m)=P(Z_1+Z_2+Z_3+\cdot \cdot \cdot + Z_n = m),
\end{equation}
\noindent where $N_c$ is a random variable denoting the total number of molecules received at the receiver nanomachine within $T_s$. Given the number of the transmitted molecules $n$, from (5) it can be seen that $N_c$ follows the binomial distribution, i.e. $N_c \thicksim \text{Bionomial}(n,p_c)$. For large $n$, distribution of $N_c$ can be approximated as Guassian distribution with the knowledge of its mean and variance as
\begin{equation}
N_c \thicksim \mathcal{N} \Big(np_c, np_c(1-p_c)\Big).
\end{equation}
\noindent Note that distribution of $N_c$  does not consider the late arriving messenger molecules from the previous symbols.

Let $N_{cp}$ be a random variable denoting the total number of messenger molecules received at the receiver nanomachine among the $n$ molecules released in the previous slot. It is evident that $N_{cp}$ is also binomially distributed as $N_c$, and can be approximated as
\begin{equation}
N_{cp} \thicksim \mathcal{N} \Big(np_{cp}, np_{cp}(1-p_{cp})\Big),
\end{equation}
\noindent where
\begin{equation}
p_{cp}=p_c  + (1-\beta)\int_{T_s}^{2T_s} f(t)dt
\end{equation}
\noindent for a given inhibition efficiency $\beta$ of the inhibitor molecules. Based on (6) and (7) the distribution of a random variable representing the number of messenger molecules that may cause inter symbol interference, denoted by $N_p$, thus can be expressed as
\begin{eqnarray}
 N_P &\thicksim & \mathcal{N} \Big(np_{cp}, np_{cp}(1-p_{cp})\Big)-\mathcal{N} \Big(np_c, np_c(1-p_c)\Big) \nonumber \\
& \thicksim & \mathcal{N} \Big(n(p_{cp}-p_c), np_{cp}(1-p_{cp})+np_c(1-p_c)\Big).
\end{eqnarray}
\noindent Utilizing distributions of $N_c$ and $N_p$ in (6) and (9), we next calculate the channel capacity and also derive the symbol error performance of the receiver nanomachine for the proposed modulation scheme.

Consider serial transmission of information symbols where binary symbols $1$ and $0$ occur with \emph{a priori} probabilities equal to $q$ and $1-q$, respectively. In such a serial transmission, probabilities of correct detection of a transmitted symbol at the receiver nanomachine (probability that $0$ is received when $0$ is transmitted and  $1$ is received when $1$ is transmitted), can be expressed as
\begin{eqnarray}
P(0,0)&=& P(0,0|S_{-1}=0)P(S_{-1}=0) \nonumber \\ &+& P(0,0|S_{-1}=1)P(S_{-1}=1) \nonumber \\
&=& (1-q) \cdot 1\cdot (1-q) + (1-q)  P(N_p < \lambda)  q \nonumber \\
&=&(1-q)^2 + q(1-q)    \Big[1 - Q(A_1)\Big] ,\ \text{and}
\end{eqnarray}
\begin{eqnarray}
P(1,1)&=& P(1,1|S_{-1}=0)P(S_{-1}=0)\nonumber \\ &+& P(1,1|S_{-1}=1)P(S_{-1}=1) \nonumber \\
&=& q P(N_c \ge \lambda)  (1-q)  + q  P(N_c+N_p \ge \lambda)  q \nonumber \\
&=& q(1-q)Q(A_2) + q^2   Q(A_3),
\end{eqnarray}
\noindent where $S_{-1}$ is a binary random variable indicating the information symbol transmitted in the previous slot, $Q(\cdot)$ is the the tail probability of the Gaussian probability distribution function,  $A_1= \frac{\lambda-n(p_{cp}-p_c)} {[np_{cp}(1-p_{cp})+np_c(1-p_c)]^{1/2}}$, $ A_2=\frac{\lambda-np_c}{[np_c(1-p_c)]^{1/2}}$ and $A_3=\frac{\lambda-np_{cp}}{[np_{cp}(1-p_{cp})+2np_c(1-p_c)]^{1/2}}$. On the other hand, probabilities of erroneous detection of a transmitted symbol at the receiver nanomachine (probability that $1$ is received when $0$ is transmitted and  $0$ is received when $1$ is transmitted), can be expressed as
\begin{eqnarray}
P(0,1)&=& P(0,1|S_{-1}=0)P(S_{-1}=0)\nonumber \\ &+& P(0,1|S_{-1}=1)P(S_{-1}=1)  \nonumber \\
&=& (1-q)\cdot 0 \cdot (1-q) +(1-q) P(N_p \ge \lambda) q \nonumber \\
&=& q(1-q)   Q(A_1), \ \text{and}
\end{eqnarray}
\begin{eqnarray}
P(1,0)&=& P(1,0|S_{-1}=0)P(S_{-1}=0)\nonumber \\ &+& P(1,0|S_{-1}=1)P(S_{-1}=1) \nonumber \\
&=&q  P(N_c < \lambda)  (1-q) + q P(N_c+N_p < \lambda) q \nonumber \\
&=& q(1-q) \Big[1-Q(A_2)\Big] +  q^2 \Big[ 1- Q(A_3)\Big].
\end{eqnarray}

\noindent Total probability that a symbol is decoded erroneously at the receiver nanomachine, denoted as $P_e$, is the sum of $P(0,1)$ and $P(1,0)$, Thus, using (12) and (13),
\begin{equation}
P_e= q(1-q)    Q(A_1) +q(1-q) \Big[1-Q(A_2)\Big] + q^2 \Big[ 1- Q(A_3)\Big].
\end{equation}

Next, we determine capacity $C$ of the considered molecular communication system, i.e., the maximum rate of transmission between the transmitter nanomachine and the receiver nanomachine. This can be calculated utilizing the seminal Shannon's formula, which defines the capacity as the maximum mutual information $I(S;R)$ between the transmitted symbol $S$ and the received  symbol $R$ as
\begin{equation}
C= \max_{\lambda}I(S;R),
\end{equation}
\noindent where
\begin{equation}
I(S;R)= \sum_{S\in{\{0,1\}}}\sum_{R\in{\{0,1\}}} P(S,R) \log_2 \frac{P(S,R)}{P(S)P(R)}.
\end{equation}

\noindent Note that $P(S,R)$  can be calculated using (10)-(13) while the marginal probabilities $P(S)$ and $P(R)$ can be obtained using $P(S)=\sum_{R\in{\{0,1\}}} P(S,R)$ and $P(R)=\sum_{S\in{\{0,1\}}} P(S,R)$.

\subsection{Numerical Results}
Fig. 3 shows mutual information between the Zebra-CSK modulated binary symbols transmitted by the transmitter nanomachine and binary symbols received by the receiver nanomachine. For all three considered cases of inhibition efficiency of the inhibitor molecules (viz. $\beta =1,\ \beta = 0.5,$ and $\beta = 0$), mutual information increases with increase in the detection threshold value until reaching its maximum but starts to fall with further increase in the detection threshold. Note that Zebra-CSK with null inhibition efficiency (i.e. $\beta = 0$) is logically equivalent to the conventional CSK. From the figure it is evident that higher mutual information can be achieved between Zebra-CSK modulated transmitted symbols and received symbols by selecting a relatively smaller detection threshold than the one that maximizes the mutual information between CSK modulated symbols and the received symbols. The higher mutual information in Zebra-CSK thus results in higher channel capacity. For instance, for the considered $n$, $d$, $q$ and $D$ values, channel capacity of Zebra-CSK is $29.85 \%$ and $37.31 \%$ higher than that of CSK when $\beta = 0.5$ and $\beta = 1$, respectively.


Fig. 4 shows symbol error performance of the receiver nanomachine while demodulating Zebra-CSK modulated binary symbols. The symbol error probability of Zebra-CSK is significantly lower than that of the conventional CSK for most of the possible detection threshold values that are smaller than the optimum detection threshold value for CSK. Interestingly, the higher the inhibition efficiency of the inhibitor molecules is, the higher will be the reduction in the symbol error probability. For instance, for the considered $n$, $d$, $q$ and $D$ values, minimum achievable symbol error probability of Zebra-CSK (i.e at optimum detection threshold) is $0.017$ when $\beta = 0.5$ and $0.00993$ when $\beta = 1$.  These are clearly $75.36 \%$ and $85.61 \%$ improvement over the minimum achievable symbol error probability of $0.069$ in the conventional CSK.

\begin{figure}[t]
\centering
\label{PE1}
\includegraphics[scale = .62]{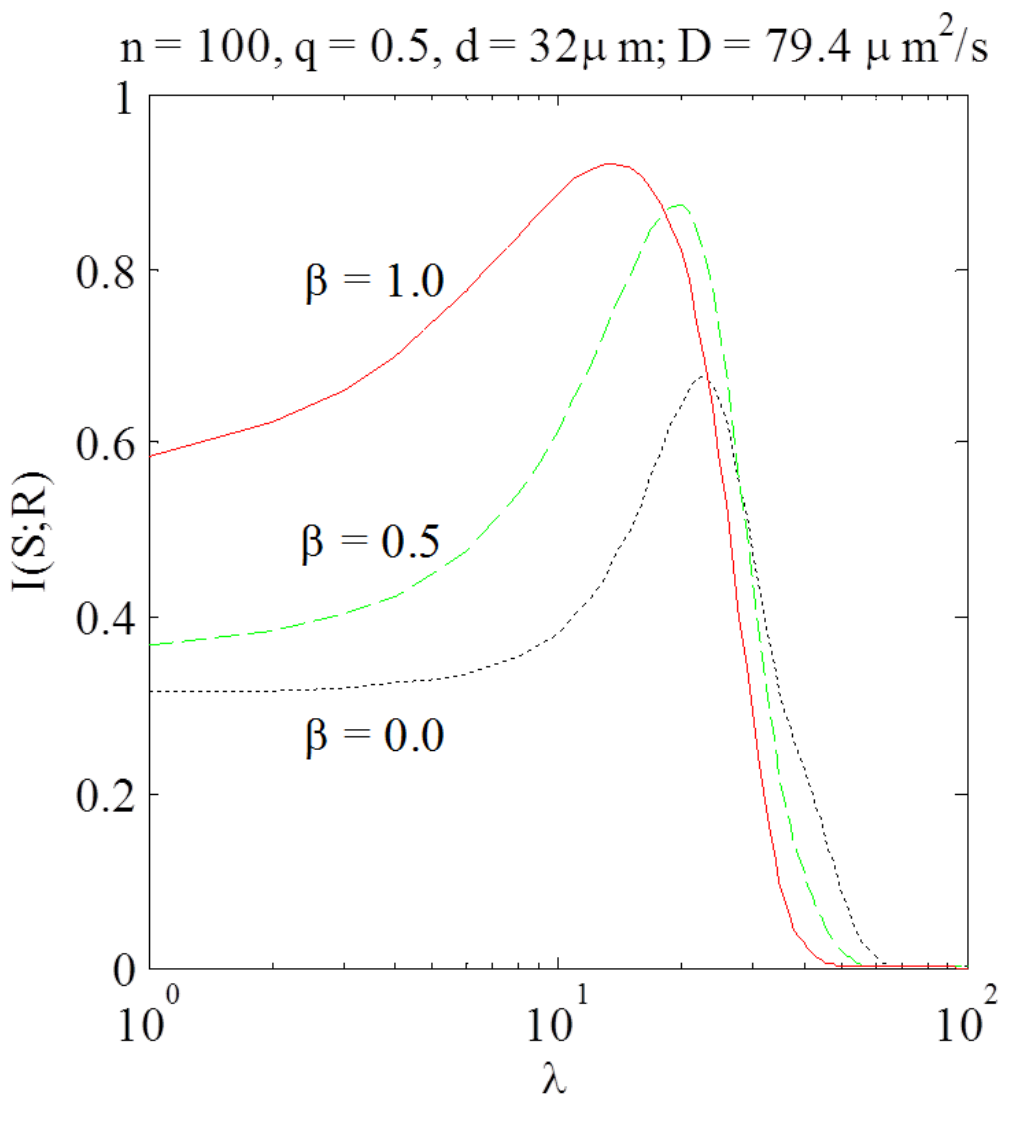}
\caption{\small Mutual information between the transmitted binary symbol $S$ and received symbol $R$ when $n=100$, $q=0.5$, $d=32\ \mu$m, and $T_s=5.9\ $s. }
\end{figure}

\begin{figure}[t]
\label{PE2}
\centering
\includegraphics[scale = .62]{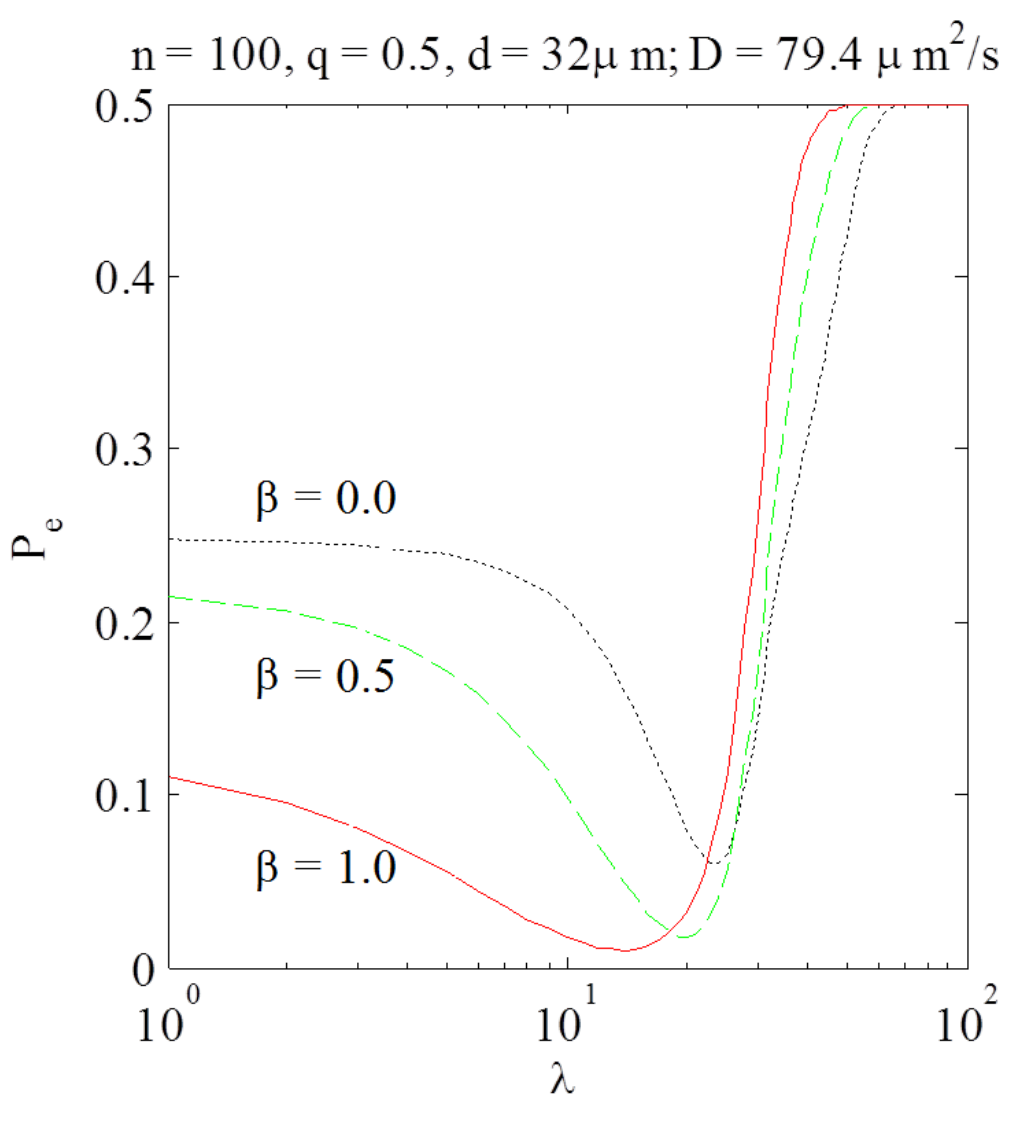}
\caption{\small Probability of symbol error at receiver nanomachine when $n=100$, $q=0.5$, $d=32\ \mu$m, and $T_s=5.9\ $s.}
\end{figure}

%


\begin{figure}[t]
\label{PE3}
\centering
\includegraphics[scale = .62]{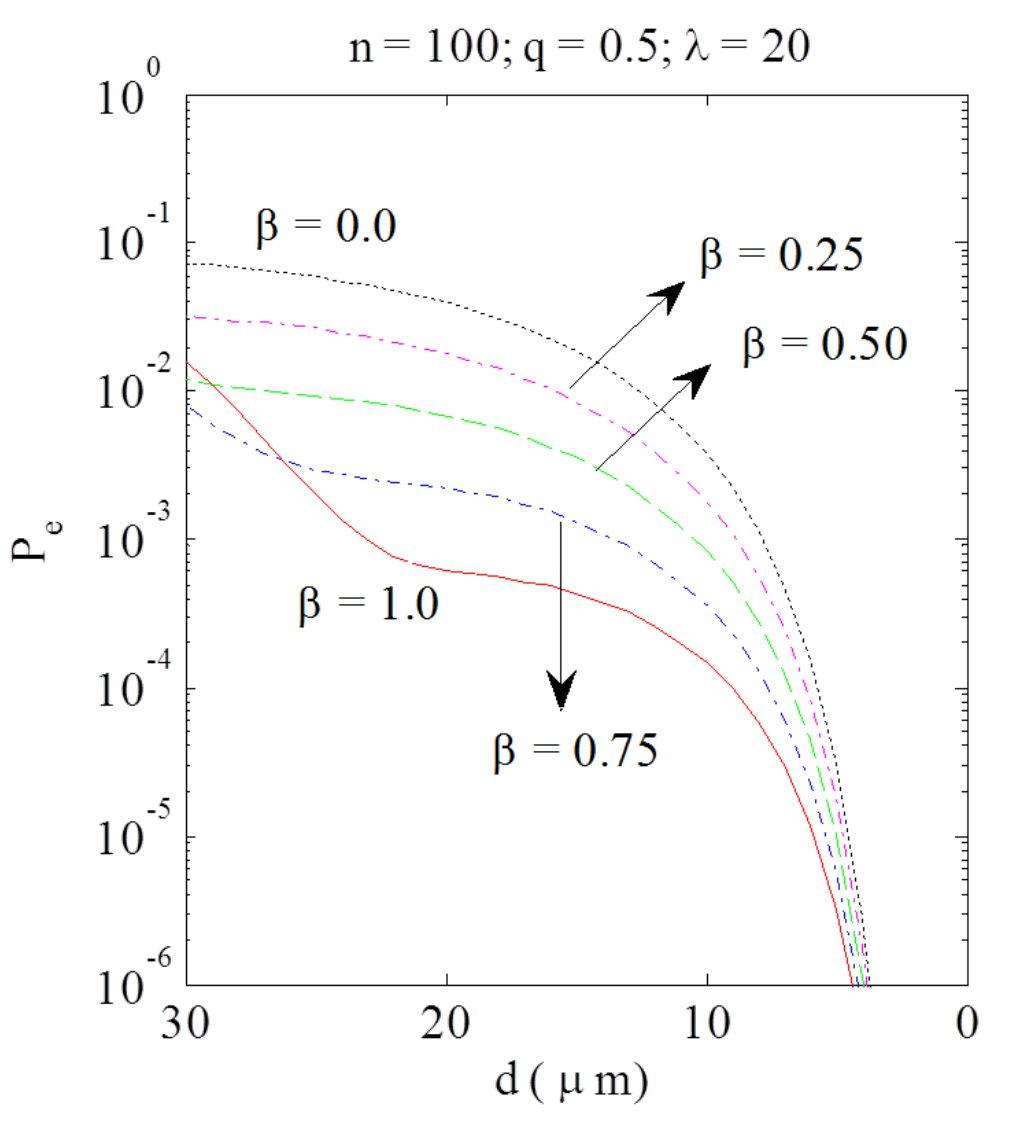}
\caption{\small Effect of distance  between transmitter and receiver nanomachine on symbol error performance when $n=100$, $q=0.5$, $T_s=5.9\ $s, and $\lambda =20$.}
\end{figure}

Fig. 5 shows symbol error performance of the receiver nanomachine located at various distance from the transmitter nanomachine. For a given detection threshold, the symbol error probability increases as the distance between transmitter and receiver nanomachine increases. Symbol error probability of Zebra-CSK is always lower than that of the conventional CSK regardless of the distance between transmitter and receiver nanomachine. It is note worthy to mention that, except for very higher inhibition efficiencies closer to $1$, symbol error performance gain of Zebra-CSK over CSK (i.e difference between the symbol error probabilities of CSK and Zebra-CSK) increases with an increase in the distance between transmitter and receiver nanomachine.

\section{Conclusion}
In this paper, we have proposed a new modulation technique (named as Zebra-CSK) for transmitting information symbols over diffusive channel in nanoscale molecular communication networks. The proposed modulation technique is robust in the sense that it reduces inter symbol interferences among subsequently transmitted information symbols. Reduction in the inter symbol interference depends on the efficiency of the molecules used to inhibit the messenger molecules. Through numerical analysis we have shown that symbol detection error probability of a receiver nanomachine is lower for demodulating Zebra-CSK modulated binary information symbols than demodulating the conventional CSK modulated binary information symbols.  As a future work,  we plan to analyze and quantify the performance of the proposed Zebra-CSK for quaternary and even higher order modulations.



\end{document}